\documentclass[12pt,epsfig]{iopart}

\usepackage{iopams}
\usepackage[figuresright]{rotating}
\usepackage{psfig}
\usepackage{epsfig}
\usepackage{floatflt}
\usepackage{graphicx}
\usepackage{dcolumn}
\usepackage{bm}

\def\Journal#1#2#3#4{{#1} {\bf #2}, #3 (#4)}

\def\NIMA{{ Nucl. Instrum. Methods} A}

\def\NPA{{ Nucl. Phys.} A}
\def\PLB{{ Phys. Lett.}  B}
\def\PRL{ Phys. Rev. Lett.}

\def\PRC{{ Phys. Rev.} C}
\def\PRD{{ Phys. Rev.} D}

\def\JPG{{ J. Phys.} G}

\begin{document}

\title{Overview of charm production at RHIC}

\author{Zhangbu Xu}

\address{Physics Department, Brookhaven National Lab., New York, NY 11973, USA}
\ead{xzb@bnl.gov}
\begin{abstract}
In this presentation, I discussed a) the charm total cross-section and
its comparisons to measurements at other beam energies and pQCD
calculations; b) the semileptonic decay of charmed hadrons and the
sensitivity of non-photonic leptons to charm quark collective flow and
freeze-out; c) semileptonic decayed electron spectrum at high
transverse momentum, its comparison to FONLL in p+p and d+Au
collisions, and heavy-quark energy loss in Au+Au collisions.
\end{abstract}


In relativistic heavy-ion collisions, charm quarks are believed to be
produced in the early stage via initial gluon fusion and their
production cross-section can be evaluated using perturbative
QCD~\cite{cacciari}.  Study of the $N_{bin}$ scaling properties of the
charm total cross-section in p+p, d+Au and Au+Au collisions can test
if heavy-flavor quarks, which are used as a probe, are produced
exclusively at the initial impact. The interactions of heavy quarks
with the medium provide a unique tool for probing the hot and dense
matter created in ultra-relativistic heavy-ion collisions at the early
times.  At RHIC energies, heavy quark energy loss~\cite{dokshitzer01},
charm quark coalescence~\cite{pbm,loic,rafelski,mclerran}, the effect
of $J/\psi$ production from charm quark coalescence on the
interpretation of possible $J/\psi$ suppression due to color
screening~\cite{matsui}, and charm flow~\cite{xu2,teaney,batsouli}
have been proposed as important tools in studying the properties of
matter created in heavy ion collisions. The last three effects depend
strongly on the charm total cross-section and spectrum at low $p_T$.

Since the beginning of RHIC operation, PHENIX and STAR collaborations
have made pioneer measurements in charm related
physics~\cite{stardAucharm,phenixAuAu,phenixpp,starcharmQM05,czhongSQM06,wangxrSQM06}
{\footnote{The overviews of charm elliptic flow and quarkonium can be
found elsewhere~\cite{shinichi,leitch}}. New measurements presented at
this conference are:
\begin{enumerate}
\item
muon spectra at forward rapidity ($1.4<|y|<2.2$) from charm
semileptonic decay by PHENIX Collaboration~\cite{wangxrSQM06}. This
enables us to study the rapidity dependence of nuclear effects of
charm production.
\item
muon spectra at low $p_T$ ($0.17<p_T<0.25$ GeV/$c$) from charm
semileptonic decay by STAR Collaboration~\cite{czhongSQM06}. This
improves the charm total cross-section measurements and better
constrains the charm spectrum for studying the charm radial flow.
\end{enumerate}
\section{Charm total cross-section}
It is difficult to directly reconstruct charmed hadrons and single
electrons from charm semileptonic decay in hadron-hadron collisions
with high precision at low $p_{T}$, where the yield accounts for a
large fraction ($\sim85\%$) of the total cross
section~\cite{stardAucharm,CDF,phenixAuAu}. The difficulties are due
to short decay distance ($c\tau\simeq100\mu m$) and large combinatorial
backgrounds in charmed hadron decay channels, and the overwhelming
photon conversions in the detector material, and $\pi^{0}$ Dalitz
decays in electron detection. Nevertheless, the charm total
cross-sections have been measured in d+Au collisions at RHIC by a
combination of the directly reconstructed low $p_T$ $D^0\rightarrow
K\pi$ and the non-photonic electron spectra~\cite{stardAucharm}, and
by non-photonic electron spectra alone at $p_T>0.8$
GeV/$c$~\cite{phenixAuAu,phenixpp}. Although the systematic and
statistical errors are large, the result indicates a much larger charm
yield than predicted by pQCD
calculations~\cite{stardAucharm,cacciari}. It was argued that results
with small renormalization (fragmentation) scales shown as solid line
in Fig.~\ref{charmpp} are not reliable calculations~\cite{cacciari}. A
new method~\cite{charmplb} was proposed to extract the charm total
cross-section by measuring muons from charmed hadron semileptonic
decay at low $p_T$ (e.g. $0.16^{<}_{\sim}p_T{}^{<}_{\sim}0.26$
GeV/$c$).  Since muons in this $p_T$ range are a very uniform sample
of the whole charmed hadron spectrum, the inferred charm total
cross-section is insensitive to the detail of the charm spectrum.
\begin{figure}
  \includegraphics[width=3.5in]{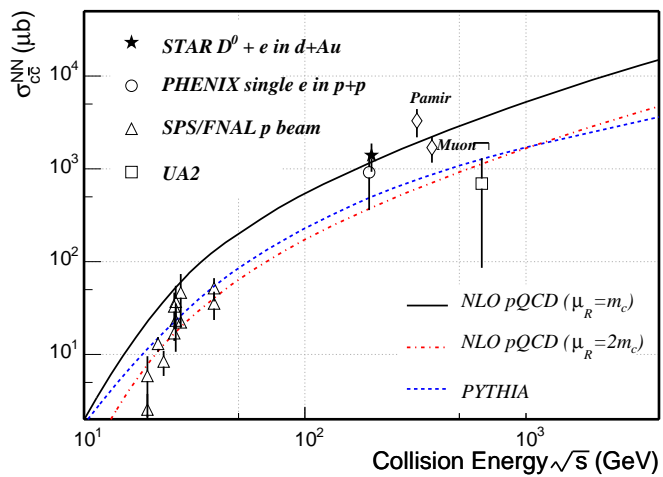}
\caption{Total $c\bar{c}$ cross-section per nucleon-nucleon collision
vs. the collision energy ($\sqrt{s_{_{NN}}}$). The dot-dashed and
dashed lines are default NLO pQCD and PYTHIA calculations. The solid
line is NLO pQCD calculations with $\mu_R$ changed from $2m_c$
(dot-dashed line) to $m_c$.}
\label{charmpp}
\end{figure}
\subsection{Are we (RHIC) consistent?}
PHENIX and STAR collaborations have used several methods and
techniques to extract total charm cross section in p+p, d+Au and Au+Au
collisions~\cite{stardAucharm,phenixAuAu,phenixpp,starcharmQM05,
czhongSQM06,wangxrSQM06}.  The total cross-section measured at
$\sqrt{s_{_{NN}}}=200$ GeV can be summarized as Fig.~\ref{charmpp} and
Fig.~\ref{charmdifference}. In general, the measurements have smaller
errors in Au+Au collisions than in p+p or d+Au collisions. The
agreements between PEHNIX and STAR are better in light systems than in
central Au+Au collisions. The non-photonic electron spectra by PHENIX
have smaller systematic errors than the corresponding STAR
measurements while STAR Collaboration have two additional measurements
from direct charmed hadron reconstruction and low-momentum muon
spectra. These reflect the strengthes of the detectors,
accordingly. There is substantial discrepancy of charm total
cross-section between those extracted from PHENIX's non-photonic
electron spectra and the combined (hadronic and semileptonic) fit
results from STAR's measurements. The results in p+p and d+Au
collisions show consistency within the errors. However, the
discrepancy is about a factor of 2 in central Au+Au collisions while
both have errors at 20\% level. Part of the discrepancy may be
explained by the different coverage of the two
experiments~\cite{charmplb}.  Measurements from PHENIX non-photonic
electron spectra cover $<15\%$ of the dN/dy of the charm
yields. Fig.~\ref{charmdifference} right panel illustrates the
possible difference between measurements at low $p_T$ and higher
$p_T$. It shows the dependence of the muon yield on power-law
parameter $\langle p_{T} \rangle$ for a fixed total charm yield
(details at Ref.~\cite{charmplb}). The yield is normalized to yields
at $\langle p_{T} \rangle=1.3$ GeV/$c$. Fig.~\ref{charmdifference}
demonstrates that over a wide range in $\langle p_{T} \rangle$, the
muon yield is within $\pm15\%$. This is in contrast to the large
variation of the electron yield integrated above $p_T$ of 1.0 GeV/$c$,
where a factor of 8 variation is seen in
Fig.~\ref{charmdifference}. On the other hand, the large discrepance
is difficult to be accounted for within reasonable range of parameters
(e.g. $\langle p_{T} \rangle$ need change from 1.4 to 0.9 GeV/$c$ for
a factor of 2 change in extrapolated total yield) when both
experimental errors are small in central Au+Au collisions. In
addition, no obvious contradiction was observed from the electron
spectra themselves and those are shown in Fig.3 and Fig.4 (details in
later sections). With upgrades by PHENIX and STAR collaborations
targeting at drastic improvement of precise secondary vertex detectors
and continuous improvements of statistics and reduction/understanding
of systematical errors, we may be able to better understand the
difference in near future~\cite{mangano}.

\begin{figure}
  \includegraphics[width=3.in]{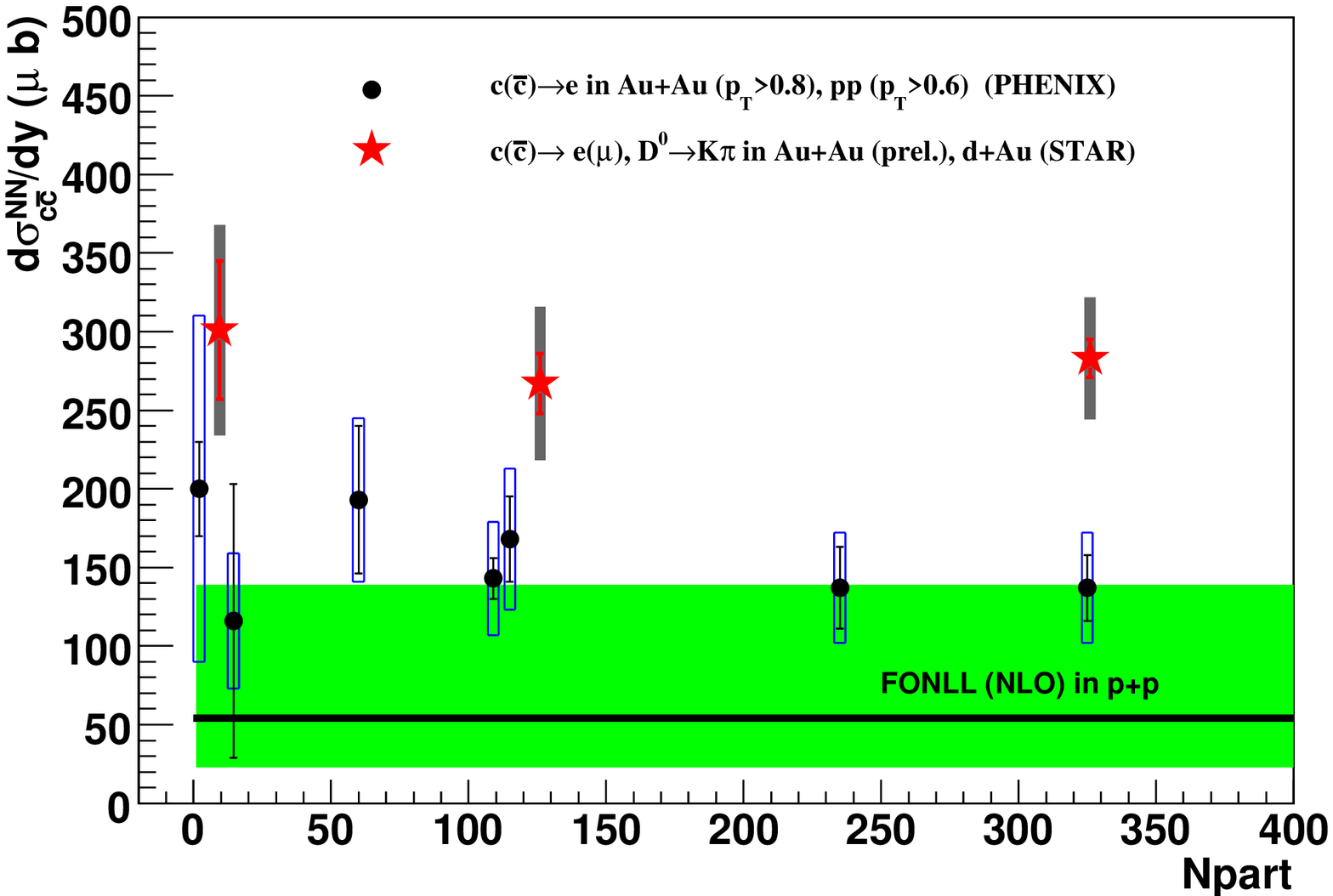}
  \includegraphics[width=3.in]{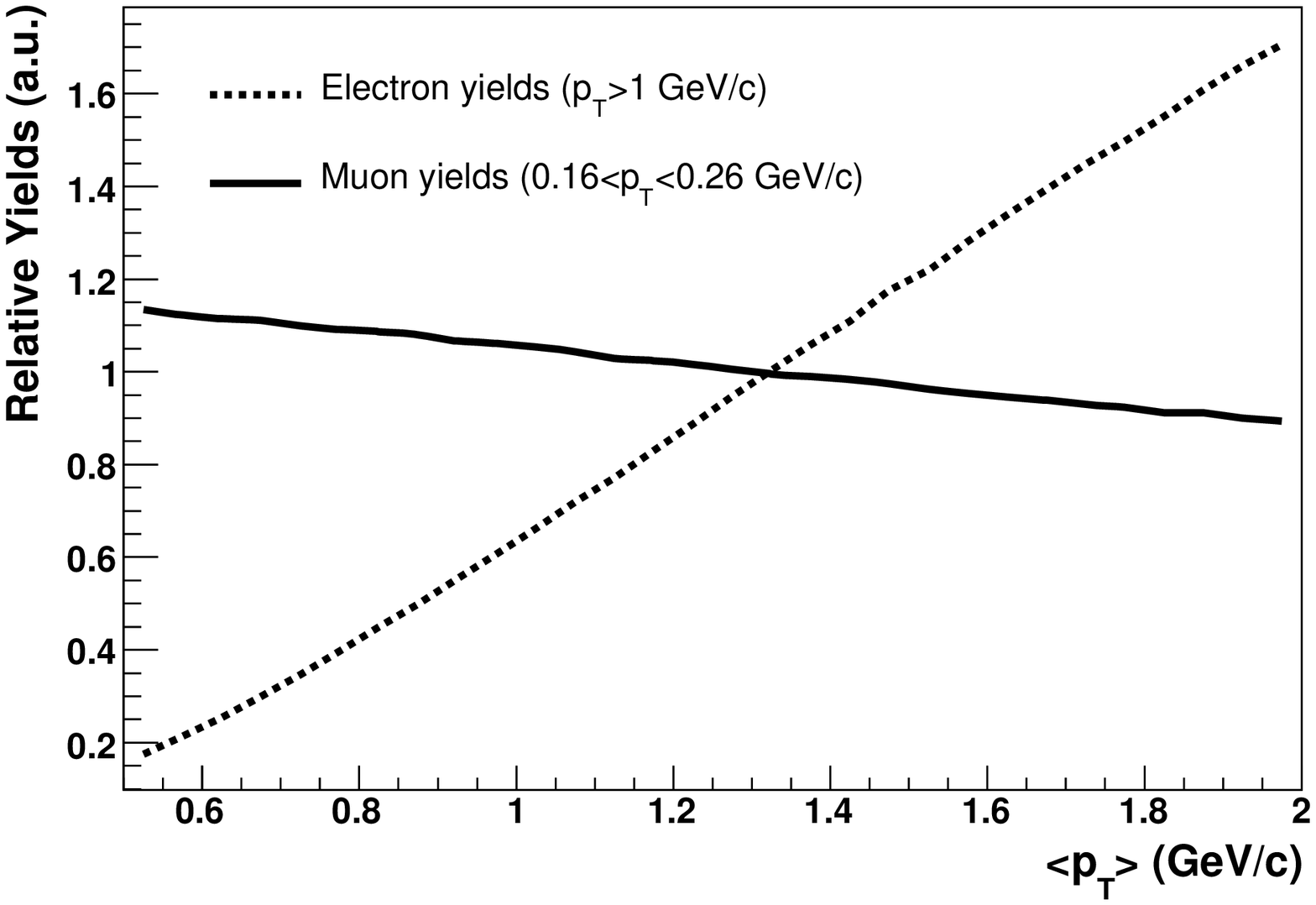}
\caption{Left panel: Differential $c\bar{c}$ cross-section per
nucleon-nucleon collision ($dN/dy$) as function of the number of
collision participant nucleons ($N_{part}$). The solid line is default
NLO pQCD and FNOLL calculations. The shaded band indicates uncertainty
of the predictions. The data points are from PHENIX and STAR
collaborations. Right panel: Lepton yields relative to the fixed total
charm cross-section as function of power-law parameters $\langle p_{T}
\rangle$ for a charmed hadron transverse momentum spectrum.  Solid
line shows muon yields with a kinematics selection $0.16<p_T<0.26$
GeV$c$ and $|y_l|<0.5$. Dashed line shows electron yields with
$p_T>1.0$ GeV/$c$.}
\label{charmdifference}
\end{figure}

\subsection{Are we alone? What's right and what's wrong?}
It has been shown that the charm cross-sections in hadron collisions
at lower energy have large errors and in many cases are
inconsistent. We examined the experimental data and how the total
cross-section was extracted. We found that in most of the cases, the
charm total cross-sections were inferred from measurements covering
small phase space using formulae to extrapolate to the full phase
space. In some cases, the functions used for extrapolation were not
consistent and resulted in large discrepancy. In addition, the
systematic error from the extrapolation has not been properly
implemented. In X. Dong's thesis~\cite{xdong}, we tabulated all the
measurements and got rid of those with large extrapolation or inferred
from correlations. Although it eliminated many measurements, those
kept show better consistency and were listed in
ref.~\cite{xdong,stardAucharm}. Now we can compare the experimental
data from low energy to high energy to the pQCD calculations. At RHIC
energies, the data points are in general above the default pQCD
predictions. The usual wisdom is that there is a factor of 2
uncertainty from pQCD calculations by varying the renormalization and
fragmentation scales within a factor of 2. We noted that cross
sections inferred from cosmic ray measurements are also much higher
than the pQCD predictions. The cosmic ray showers required large charm
total cross-section to account for the large muon yields and
electromagnetic showers within the cosmic ray shower with incident
nucleon energy at a few teens of TeV. On the other hand, the results
at lower energy are consistent with pQCD calculations as shown in
Fig.~\ref{charmpp}, and charmed hadrons at $p_T>5$ GeV/$c$ at Tevtron
is within a factor of 2 above pQCD calculation~\cite{CDF}. We have
also compared the non-photonic electron spectra from ISR energies to
RHIC and to Tevtron energies. We found that there is no obvious
inconsistency among the electron spectra. The direct $D^{0}$
measurement for $p_T>5$ GeV/$c$ in $p+\bar{p}$ collisions by CDF
Collaboration was fitted to a power-law function and the semileptonic
decayed electron spectrum was from the extrapolated spectrum. Detail
of this study has been shown in Ref.~\cite{xdong,xuISMD}.

\section{Are leptonic spectra sensitive to charm flow?}
We propose a new method to extract the charm total cross-section by
measuring muons from charmed hadron semileptonic decay at low $p_T$
(e.g. $0.16^{<}_{\sim}p_T{}^{<}_{\sim}0.26$ GeV/$c$).  Since muons in
this $p_T$ range are a very uniform sample of the whole charmed hadron
spectrum, the inferred charm total cross-section is insensitive to the
detail of the charm spectrum. Once the cross-section is determined,
the electron spectrum at higher $p_T$ can be used to sensitively infer
the charmed hadron spectral shape.  Meanwhile, we survey the form
factors used in charm semileptonic decays generated from Particle Data
Group~\cite{pdgcharmff}, in the PYTHIA event
generator~\cite{stardAucharm,phenixAuAu,phenixpp,pythia}, by pQCD
predictions~\cite{cacciari} and from the CLEO inclusive
measurement~\cite{cleocetalk}.  We find that the lepton spectra from
these different form factors can be different by a factor of 1.5.
\subsection{Form factors}
The spectrum generated by the PDG is according to the form factor of
charmed meson decays to pseudoscalar $K+l+\nu$, vector meson
$K^*+l+\nu$ and non-resonance $(K\pi)+l+\nu$ where the $K^*$ mass is
used for the $(K\pi)$ system. Since PYTHIA uses a simplified vector
meson decay form factor~\cite{pythia}, it tends to produce a softer
electron spectrum. Both the parameterization by
Cacciari~\cite{cacciari} and formulae from the PDG agree with CLEO's
electron spectrum~\cite{cleocetalk}. In addition, we also find that
although the charmed mesons ($D^{\pm}$ and $D^{0}$) from $\Psi$(3770)
decay have a momentum of 244 MeV/c only and without correction of
final state radiation~\cite{cleocetalk}, it affects slightly its
subsequent electron spectrum.
\subsection{Freeze-out}
The semileptonic decay greatly smears the spectrum and reduces the
difference between the different spectrum shapes. However, it is clear
that a reasonably realistic blast-wave parameterization of charmed
mesons in Au+Au collisions is very different from that in d+Au
collisions. There is also a significant difference between spectra
with different flow (blast-wave function) parameters at $0.5<p_T<1.5$
GeV/$c$. Fig.6 in Ref.~\cite{charmplb} shows that there is a factor of
3 difference at $p_T=1.5$ GeV/$c$ between late freeze-out
($T_{fo}=100$ MeV, $\beta_{max}=0.9$)~\cite{olga} and early freeze-out
($T_{fo}=160$ MeV, $\beta_{max}=0.6$)~\cite{multistrange}. Current
measurements of non-photonic electron spectra and direct charm spectra
seem to be consistent with a decreasing trend even at $p_T\simeq1.0$
GeV/$c$~\cite{czhongSQM06,starcharmQM05}, which is likely due to
multiple collisions and thermalization at low $p_T$ with early
freeze-out~\cite{teaney} and not due to pQCD energy loss. However,
since the overall normalization is not known, an early freeze-out
scenario can be interpreted as suppression of the charm total
cross-section as well. This ambiguity can be resolved by a measurement
of the total cross-section~\cite{charmplb}. In fact, with the
measurements by PHENIX and STAR collaborations presented in this
conference, it was demonstrated that we are able to obtain freeze-out
parameters based on a blast-wave assumption with reasonable
errors~\cite{czhongSQM06} (also shown as solid line in
Fig.4). However, the errors on the current measurements are large in
this $p_T$ range. We advocate improving the measurements of electrons
in this $p_T$ range to assess if charm thermalizes in the
medium~\cite{xu2}.
\section{Does it make senses: Color and Flavor?}
Recently, the measurements of high $p_T$ electrons from heavy-flavor
semileptonic decays have posed challenges to our understanding of
partonic energy loss in the
medium~\cite{phenixAuAu,czhongSQM06,starcharmQM05,miklos,xdong}.
\subsection{p+p/FONLL is 5.5 at high $p_T$?}
It has been perceived that pQCD can reasonably calculate charm
production because its quark mass is much larger than the
non-perturbative scale. However, we have shown that the total cross
section seems to be above pQCD calculations in several cases. On the
other hand, the agreement should be better at higher $p_T$.
Fig.~\ref{charmpp2FONLL} shows the measurements of non-photonic
electron spectra and muon spectra in p+p collisions divided by the
First-Order-Next-to-Leading-Log (FONLL) calculations~\cite{cacciari}
as function of $p_T$. The pQCD calculations have a factor of 2
uncertainty. Although the experimental data have large errors at low
$p_T$, it is in general larger than the prediction. In particular, the
STAR data from Electromagnetic calorimeter and the PHENIX data from
muon detector are a factor of 5 above the FONLL calculations. This
presents a puzzling issue since charged hadron and
pions~\cite{rhicwhitepapers,highptppi} at high $p_T$ can be quite well
reproduced by NLO pQCD calculations.
\begin{figure}
  \includegraphics[width=3.5in]{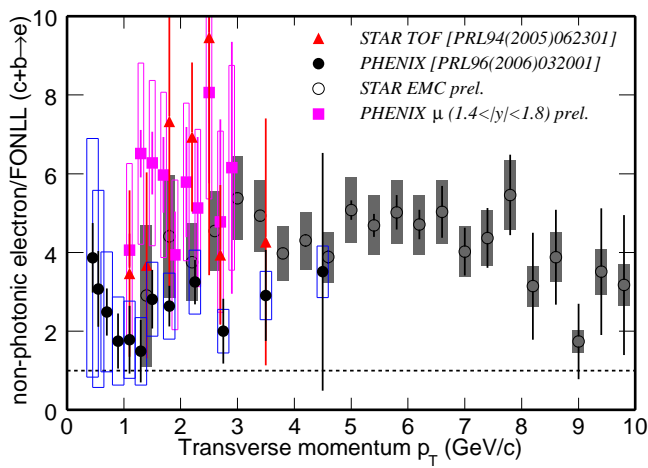}
\caption{Ratio of non-photonic electrons to semileptonic decayed
electrons from charmed and bottomed hadrons in FONLL as function of
$p_T$. The data points are from PHENIX and STAR collaborations.}
\label{charmpp2FONLL}
\end{figure}
\subsection{Color and flavor dependence of energy loss}
Let us ignore the discrepancy between data and pQCD in p+p collisions
and study the nuclear modification function ($R_{AA}$) in Au+Au
collisions. PHENIX and STAR collaborations has shown that $R_{AA}$ of
non-photonic electron (presumedly from charm and bottom decays) is
much smaller than 1 for $p_T{}^{>}_{\sim}2$ GeV/$c$ shown in
Fig.~\ref{RAAe}. In fact, it is very similar to that of pions with
$R_{AA}\simeq0.2$.  Jet-quenching models incorporating collisional and
radiation energy loss can not account for such large
suppression~\cite{miklos}. This may be due to the large contribution
of bottomed hadrons to the non-photonic electron in the model
calculations, which assume that charm and bottom production scale the
same way from pQCD to match data in p+p collisions~\cite{liuko}.
\begin{figure}
  \includegraphics[width=3.5in]{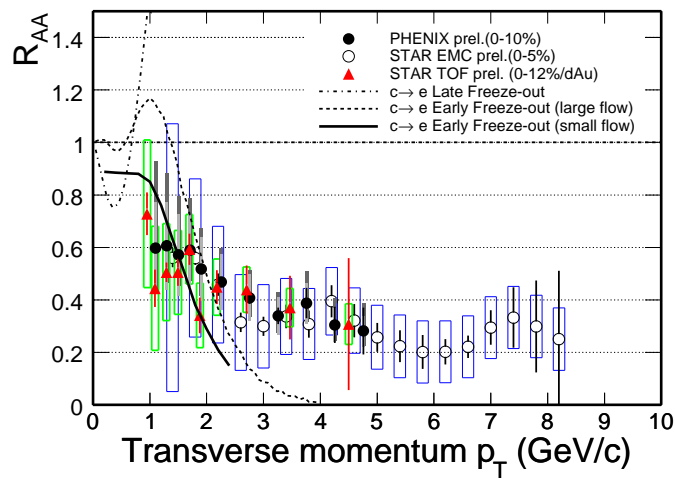}
\caption{The nuclear modifiction function ($R_{AA}$)of non-photonic
electron spectra vs. transverse momentum. Curves show different
freeze-out assumptions in a blast-wave model. The data points are from
PHENIX and STAR collaborations.}
\label{RAAe}
\end{figure}

If we look at the $R_{AA}$ at quark level for gluons, light quarks,
and charm quarks, it is obvious that the difference between light
quarks and charm quarks is quite small (much less than a factor of two
at low $p_T$ and similar at high $p_T$) calculated by
Ref.~\cite{miklos}. The largest difference is between light quarks and
gluons. This means that the deadcone effect of heavy quarks traversing
the medium has much smaller effect on the nuclear modification factor
than the color-charge factor of 9/4 between gluons and quarks. To
study this, it is important to find differential experimental probes
sensitive to gluon and light quark energy loss. It has been
proposed~\cite{wang98} that anti-protons at high $p_T$ (10 GeV/$c$) in
A+A collisions are mainly from gluon fragmentation while pions are
dominated by light quark fragmentation. By measuring $R_{AA}$,
$\bar{p}/p$ and $\bar{p}/\pi$ ratios, we will be able to access the
difference between gluon and quark energy loss. This was done by STAR
Collaboration in a recent publication~\cite{highptppi} using the
relativistic rise of ionization energy loss in TPC to separate protons
and pions~\cite{pidNIMA}. The results show that there is no difference
between proton and pion $R_{AA}$ (or $R_{CP}$) and little centrality
dependence of $\bar{p}/\pi$ ratio has been found~\cite{highptppi}.

We now witness a set of data showing that the nuclear modification
factors are the same among hadrons that are presumedly fragments from
separated gluons, light quarks and heavy quarks. This apparently
contradicts the prediction from jet quenching models, which
successfully explain the light hadron production at high $p_T$ and
dijet correlations~\cite{rhicwhitepapers}. It doesn't necessarily mean
that the general framework of jet quenching is invalid. This may imply
that interesting phenomena are present besides the general idea of
energy loss of energetic partons traversing dense medium.

Experimentally, we need to measure the charm total cross-section
with energy scan to map out its energy dependence and help constrain
the pQCD calculations and charm coalescence into $J/\Psi$. We also
will be able to study radial and elliptic flows of heavy flavors
with upgrades by PHENIX and STAR collaborations. We need to
separately measure the nuclear modification factors of charmed and
bottomed hadrons and possible heavy-quark tagged jets to study the
charm and bottom quark energy loss.

The author thanks Drs. S. Blyth, P. Braun-Munzinger, X. Dong,
J. Dunlop, H.D. Liu, B. Mohanty, T. Ullrich, A. Suiade, X.R. Wang,
N. Xu, H.B. Zhang, Y.F. Zhang, and C. Zhong for valuable discussions
and for providing the preliminary collaboration data. This work was
supported in part by the HENP Divisions of the Office of Science of
the U.S. DOE and in part by a DOE Early Career Award and the
Presidential Early Career Award for Scientists and Engineers.

\end{document}